\documentclass[12pt]{iopart}

\usepackage[english]{babel}
\usepackage{%
  epsf,
  textcomp,
  units,
}%

\newcommand{\degree}{\ensuremath{\textrm{\textdegree}}}
\newcommand{\abs}[1]{\ensuremath{\vert#1\vert}}%
\newcommand{\isi}[2]{\ensuremath{\Delta\left(\alpha_{#1},\alpha_{#2}\right)}}%
\newcommand{\brks}[1]{\ensuremath{\left\lbrack#1\right\rbrack}}%
\newcommand{\brcs}[1]{\ensuremath{\left\lbrace#1\right\rbrace}}%
\newcommand{\prns}[1]{\ensuremath{\left(#1\right)}}%

\begin{document}

\title[Biosonar classification]{A computational theory for the
  classification of natural biosonar targets based on a spike code}
\author{Rolf M{\"u}ller}
\address{Dept. Animal Physiology, T{\"u}bingen University, Morgenstelle 28, D-72076 T{\"u}bingen, Germany}
\ead{rolf.mueller@uni-tuebingen.de}

\begin{abstract}
  A computational theory for classification of natural biosonar
  targets is developed based on the properties of an example stimulus
  ensemble. An extensive set of echoes ($84\,800$) from four different
  foliages was transcribed into a spike code using a parsimonious
  model (linear filtering, half-wave rectification, thresholding). The
  spike code is assumed to consist of time differences (interspike
  intervals) between threshold crossings. Among the elementary
  interspike intervals flanked by exceedances of adjacent thresholds,
  a few intervals triggered by disjoint half-cycles of the carrier
  oscillation stand out in terms of resolvability, visibility across
  resolution scales and a simple stochastic structure
  (uncorrelatedness). They are therefore argued to be a stochastic
  analogue to edges in vision. A three-dimensional feature vector
  representing these interspike intervals sustained a reliable target
  classification performance (0.06\% classification error) in a
  sequential probability ratio test, which models sequential
  processing of echo trains by biological sonar systems.  The
  dimensions of the representation are the first moments of duration
  and amplitude location of these interspike intervals as well as
  their number. All three quantities are readily reconciled with known
  principles of neural signal representation, since they correspond to
  the center of gravity of excitation on a neural map and the total
  amount of excitation.
\end{abstract}

\submitto{\NET}

\maketitle

\section{Problem}\label{sec:problem}

This work explores a computational theory for a set of biosonar tasks
faced by bats. Based on an extensive set of real world echo data, it
develops and explores a parsimonious solution for a well-defined, yet
widely useful set of sensing problems posed by extended, multi-faceted
sonar targets. In particular, classification of foliages from
different species of deciduous trees is performed. Such foliages are
examples of ubiquitous, natural sonar targets in the habitats of many
bat species. The ability to classify them is immediately relevant to
biological tasks like landmark identification or habitat evaluation
(e.g., based on a probability estimate for the presence of a certain
prey) in general.  Furthermore, in any other estimation task where the
informative signal properties depend on foliage class, a hypothesis
for the latter could be employed to enhance performance.  Examples of
other related biological tasks likely to be performed to some extent
by bats could be related to obtaining information about the convex
hull of an extended target or finding passageways (e.g., in collision
avoidance, contour following or path planning).\par

Multi-faceted targets, which place moderate to large numbers of
reflectors in the sonar beam, pose a special challenge for sonar
systems limited to sparse spatial sampling with only two receivers:
Echoes received by each ear are superpositions of contributions from
all reflectors within the beam (moderate facet numbers would be on the
order of $10$, large numbers on the order of $10^2$ to $10^4$).
Reconstruction of target geometry/reflector location would require
both deconvolution (bats use chirping sonar pulses) as well as
estimating reflector placement from a collection of integrals over
prolate spheroidal surfaces. The second step in particular - besides
relying on simplifying assumptions~\cite{Kak-A2001} not necessarily
met in natural biosonar targets - will remain an ill-posed problem
until a sufficiently large number of such integrals has been gathered.
The behavioral patterns seen in bats may not leave enough room for
this prior to the time when a class estimate is due.  Besides the
issue of possible intractability under such constraints, a parsimony
argument stands against reflection-tomographic solutions as a model
for biosonar function in these tasks: Position, orientation and shape
of individual reflectors in a foliage are not immediately relevant to
the behavioral goals of the animal and therefore reconstruction of
these target features would be a detour into yet another
representation from which the relevant variables (identity of a
landmark, collision risk, presence and location of a passageway, etc.)
would still have to be estimated.  Parsimonious models for biosonar
sensing should neither recover irrelevant detail about a target
explicitly nor should they rely on intermediate representations which
contain an excessive amount of such detail.\par

If the geometry of a target is not known, it is impossible to predict
the waveform or other individual properties of subsequent echoes
received from it at different viewing positions. In this sense, the
echoes from foliages have to be viewed as realizations of random
processes, despite their origin in a deterministic reflection process.
Consequently, the particular problem at hand here is to classify
natural targets (foliages) based on random input signals, where the
individual waveforms will in general not contain any deterministic
patterns beyond the sonar pulse used to generate
them~\cite{Mueller-R2000a}. The computational theory presented here
deals with performing this task based on a simple spike code. In
contrast to a previous attempt at solving this
problem~\cite{Kuc-R2001}, the present work is based on a thorough
characterization of the stimulus ensemble, explores the fundamental
nature of the employed coding scheme and evaluates the performance of
the proposed estimator quantitatively.\par

Since bats emit trains of pulses, this evaluation of the proposed
estimator will take the form of an m-ary sequential probability ratio
test. In this way, it will be explored to what extent bats could make
use of the sequential information that they receive in their pulse
trains.

\section{Aim of the paper}\label{sec:aim}

The work presented here solves the problem outlined above based on
features derived from a parsimonious model of the signal
representation formed in the auditory system. This serves a dual
purpose: First, it helps to outline a solution space for
classification of natural targets into which the specific solutions
adopted by bats must fall. Second, it employs known functional
principles from biology as a means to discover good solutions to
fundamental problems with wider relevance to technical applications.

Like most mobile animals, bats possess navigation skills and the
ability to make habitat choices. This work demonstrates echo features
which have the necessary explanatory power to qualify as a tangible
hypothesis for the basis of these skills. The fact that these features
have been proven effective with realistic, physical data, makes them
excellent candidates for testing their actual use by bats in
behavioral and neurophysiological experiments. Since the work
presented here is a computational theory, it is concerned with
principle feasibility only and does not consider functional properties
of the auditory system with no specific relevance to the particular
problem at hand.

The specific biosonar sensing problem under consideration can be taken
as an example of a wider class of random signal classification
problems. Related problems arise in technical applications, e.g.,
biomedical ultrasound diagnosis or channel estimation for wireless
communication links. Bats provide an existence proof for the solutions
to problems associated with the tasks the animals face and hence offer
a convenient access route to more general solutions of possible
technological relevance.

\section{Approach}\label{sec:approach}

The approach taken here is to employ a biomimetic sonar observer which
selectively replicates those fundamental functional properties of its
biological paragon that are relevant to the particular problem at
hand. The biomimetic observer is used to collect large echo data sets
from extended, natural targets over a realistic range of viewing
positions. In this way, the natural variability can be exhausted for
these particular examples and statistical characterizations of the
stimulus ensemble can be obtained with sufficient confidence even if
they require a large number of data points (e.g., non-parametric
estimates of multivariate probability density functions). In the
present work, the stimulus ensemble is characterized at the level of
spike code features.  The spike code features are the result of
processing the experimental stimulus ensemble with a parsimonious
spike generation model.  Consequently, the identified features are
salient under a minimum number of assumptions as well.

\subsection{Biomimetic sonar system and data}\label{sec:data}

Hedges of four deciduous tree species, sycamore (\textit{Platanus
  hybrida}), linden (\textit{Tilia cordata}), field maple
(\textit{Acer campestre}), and hornbeam (\textit{Carpinus betulus}),
were constructed from large individual branches. These targets
extended between 2 and \unit[3]{m} in width, $\sim\unit[2]{m}$ in
height and between 1.6 and \unit[2]{m} in depth.  Each hedge was
composed of 3 to 8 individual branches, which were arranged to fill
the given volume in a semi-natural fashion. The targets were
considerably larger along every dimension than those employed by
\cite{Mckerrow-P2001} in a study on foliage classification with CWFM
sonar, where target depth appeared to be less than \unit[45]{cm},
making individual plant shape a likely determinant of the observed
features. Just like in a natural forest, where individual trees are
almost certain to extend beyond the volume which can be illuminated by
an individual sonar pulse, this was not the case here.\par

\begin{figure}[htbp]
  \begin{center}
    \epsfbox{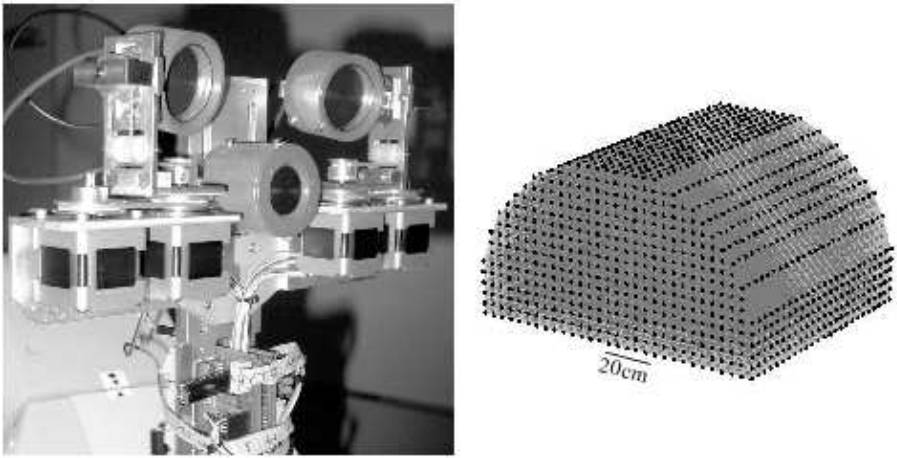}
  \end{center}
  \caption{\label{fig:methods}Experimental setup and spatial sampling
    paradigm. Left: biomimetic sonar head; right: spatial arrangement
    of points (black dots) at which echoes were sampled. The gray
    surfaces represent the convex hull of all sampling positions. The
    target hedge was positioned opposite to the frontal face of the
    sampled volume.}
\end{figure}
The targets were scanned in three dimensions with a biomimetic sonar
head (see \fref{fig:methods} left) mounted on a humanoid robot
arm.  The sonar head consisted of three electrostatic transducers, one
for emission (Polaroid 7000) and two for reception (Polaroid 600). The
receivers were positioned \unit[12.5]{cm} apart (measured between
aperture centers); the emitter was placed halfway between and
\unit[4.5]{cm} below the two receivers.  The head was moved within a
work envelope of \unit[116]{cm} width, \unit[64]{cm} height and
\unit[96]{cm} depth (perpendicular to the hedge). Since the two upper
edges of the work envelope perpendicular to the target were rounded
due to lack of reachability, the entire scanned volume was
$\sim\unit[0.6]{m^3}$ (as opposed to $\sim\unit[0.71]{m^3}$ for a
cuboid of the given edge lengths). The minimum target range within
this work envelope was $\sim\unit[1\,\textrm{to}\,1.3]{m}$.\par

The directivity of the employed electrostatic transducers is modeled
well by an (unbaffled) piston~\cite{Bozma-O1991,Morse-PM1986}, which
is also in fairly good agreement with data from at least two bat
species~\cite{Hartley-DJ1989,Coles-RB1989}. The first-null beamwidth
is \unit[40]{\degree} for the emitter and \unit[30]{\degree} for the
receivers.  These beams correspond to sonar footprint diameters of
$\sim\unit[73]{cm}$ and $\sim\unit[54]{cm}$ in \unit[1]{m} distance,
respectively (assuming normal incidence). While both emitter and
receivers were always oriented towards the hedge, this did not
guarantee normal incidence since the local orientation of the hedge
surface varied and the data can be expected to represent a wider range
of grazing angles.\par

\begin{figure}[htbp]
  \begin{center}
    \epsfbox{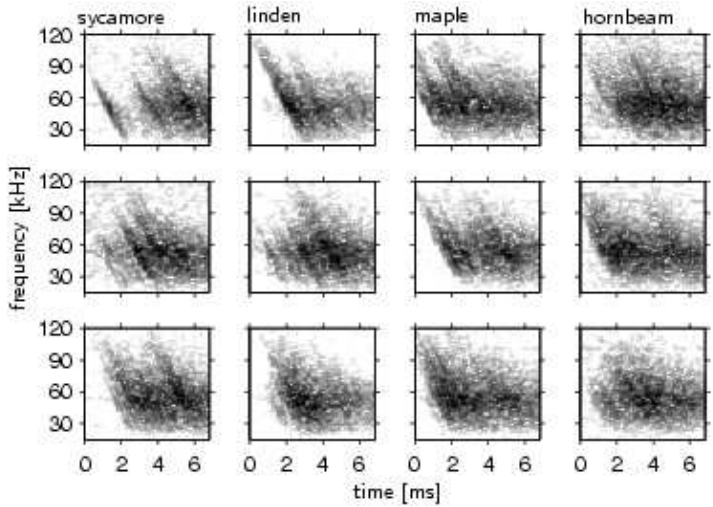}
  \end{center}
  \caption{\label{fig:waveform_examples} Spectrograms of example
    echoes. For each of the four target classes (columns), three
    examples (rows) are shown. The spectrograms were computed using a
    Hamming window spanning 512 samples, windows were spaced for an
    overlap of $3/4$ of their width. All spectrograms were normalized
    for equal maximum power and the dynamic range was restricted to
    \unit[50]{dB}.}
\end{figure}

\begin{figure}[htbp]
  \begin{center}
    \epsfbox{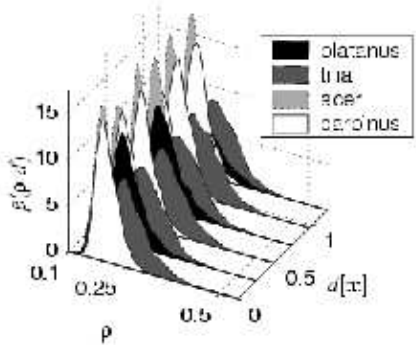}
  \end{center}
  \caption{\label{fig:echo_xcorr_d}Conditional probability density
    function estimates $\hat{p}(\rho|d)$ of the maximum correlation
    coefficient $\rho$ (over all lags, see~\eref{eq:corr_def}) between
    echoes conditioned upon the distance $d$ between the recording
    positions.  Density estimates were based on a random sample of
    $1\,000$ echo pairs per distance bin and foliage class and were
    obtained with normal kernels (smoothing bandwidths between $0.004$
    and $0.012$), the asymptotic mean integrated squared error (AMISE,
    the first order term in a series expansion of the mean integrated
    square error,~\cite{Scott-DW1992}) ranges between $0.018$ and
    $0.067$.}
\end{figure}

The volume enclosed by the work envelope was sampled every
\unit[4]{cm} along the width, height and depths dimensions (see
\fref{fig:methods} right), resulting in $10\,600$ positions and a
total of $21\,200$ echoes received at the two ``ears'' for each
target. The total data set size for all four targets is therefore
$84\,800$. Echo waveforms were digitized with \unit[1]{MHz} sampling
rate and \unit[12]{Bit} resolution. Spectrograms of example echoes are
shown in \fref{fig:waveform_examples}.
Regardless of distance between recording positions, all echoes in the
data set showed very low correlations determined over all possible
lags ($\tau$) as
\begin{equation}
  \label{eq:corr_def}
  \hat{\rho}=\frac{\max\{|\hat{C}_{xy}(\tau)|\}}{\sqrt{\hat{E}_x\hat{E}_y}},
\end{equation}
where $C_{xy}$ is the biased estimate of the cross-covariance between
the two echoes~\cite{Jenkins-GM1968} and $\hat{E}_x,\hat{E}_y$ are
estimates of their respective energies (\Fref{fig:echo_xcorr_d}).
A thorough reshuffling of weights for each reflector due to the
directivities of reflectors and transducers is the likely cause for
these small correlation distances, which do not exceed the sampling
distance chosen here for any correlation value of practical relevance.

\subsection{Biological signal processing model}\label{sec:model}

The signal processing model used for characterizing the stimulus
ensemble at a spike code level consists of two stages: preprocessing
and spike generation. Both stages were simplified to reflect only
essential signal processing steps.

\begin{figure}[htbp]
  \begin{center}
    \epsfbox{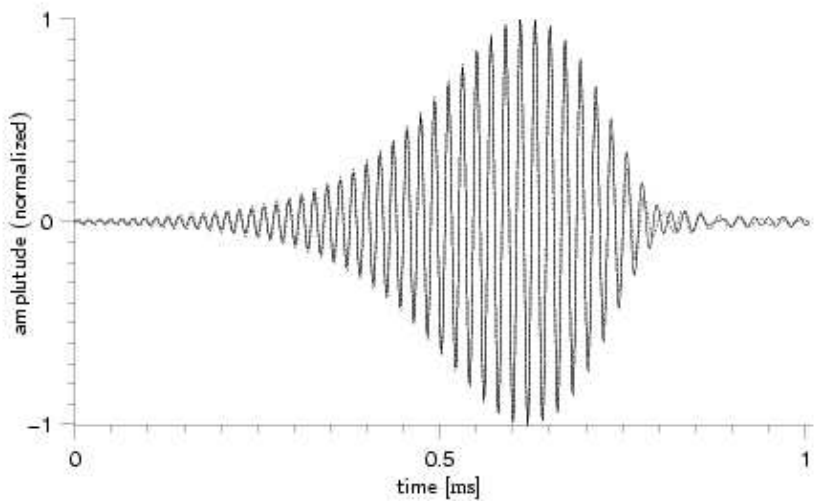}
  \end{center}
  \caption{\label{fig:chirplet}Measured combined impulse response
    (chirplet) for sonar pulse, emitter, receiver, and auditory
    bandpass filter model (\full). The auditory bandpass model
    parameters $f_c=\unit[50]{kHz}$ and $Q_{\unit[-3]{dB}}=10$ are
    used throughout the reported work and result in a \unit[-3]{dB}
    impulse response duration of $\sim\unit[260]{\textrm{\textmu}s}$.
    The actual data of the combined impulse response was collected by
    directing the sonar head at a plane in $\sim\unit[1.8]{m}$
    distance; in the graph, it is compared to a simplified model which
    omits the transducer transfer functions (\dashed).}
\end{figure}

In the preprocessing stage, the reflector sequence (impulse response)
of the target was filtered by four bandpass filters in series: the
emitted pulse, the transfer functions of emitter and receiver, as well
as an auditory bandpass filter model. The emitted pulse was a linearly
frequency modulated chirp sweeping across almost the entire passband
of the transducers (from \unit[120]{kHz} to \unit[20]{kHz}) in
\unit[3]{ms}. As the first major simplification introduced here, only
a single bandpass channel in the auditory representation of this
wideband signal is considered. A 4-th order gammatone filter with
center frequency $f_c$ and \unit[-3]{dB} quality $Q$ (ratio of $f_c$
and the filter bandwidth at \unit[-3]{dB}) was used as an accepted
standard~\cite{Slaney-M1993} for modeling auditory filters, although
the specific shape of the transfer function is of little relevance to
the features that this work focuses on. The combined effect of all
linear signal processing stages can be described as filtering the
reflector sequence with a chirplet, which is the result of convolving
all four impulse responses (\Fref{fig:chirplet}).  Since the passbands
of the transducer transfer functions are broad compared to that of the
auditory filter model, their effect on the combined impulse response
is negligible for any particular auditory bandpass channel observed in
isolation. Depending on the width of the channel's passband, the
frequency sweep of the pulse will also be negligible, resulting in the
combined impulse response being approximately a wavelet of constant
carrier frequency.
Preprocessing was completed by an approximate envelope extraction
performed as half-wave rectification and subsequent lowpass
filtering~\cite{Dau-T1996}. To the extent to which this procedure
provides for an undistorted demodulation of the
signal~\cite{Mueller-R2000b}, the effect of the lowpass filter is
equivalent to a further increase in the quality of the original
bandpass filter. The employed lowpass filter was a 1st order recursive
lowpass filter (``leaky integrator'') with time constant $\tau$.
Altogether, the simplified preprocessing model is described by a
parameter triplet of $f_c, Q$ and $\tau$. Throughout this work, the
center frequency of the auditory model filter was set to
\unit[50]{kHz}, close to the maximum of the transducer transfer
function.  The chosen filter quality of 10 at \unit[-3]{dB} is
approximately commensurate with findings from some nuclei in the lower
auditory brainstem~\cite{Haplea-1994}. Although integration times in
bats have been probed with both psychoacoustic and physiological
methods~\cite{Weissenbacher-P2002}, it is difficult to obtain an
estimate for the parameter $\tau$ of the present model from these
experimental results. Therefore, the entire range of plausible values
was explored here (from $\tau=0$ to \unit[10]{ms}, s. below) and the
cases $\tau=0$ (no integration) and $\tau=\unit[3]{ms}$ (significant
smoothing, yet far from perfect integration) are shown throughout as
examples to assess the influence of further integration/narrowing of
the passband on all reported results. In evaluating model performance
for target classification (see \sref{sec:classification}), a
broader sample of model parameter combinations ($f_c,Q,\tau$) was used
($f_c=\brcs{\unit[40]{kHz};\unit[45]{kHz};\unit[50]{kHz};\unit[55]{kHz};\unit[60]{kHz};\unit[65]{kHz}},
Q=\brcs{10;15;20;25;30;35},\tau=\brcs{0;\unit[1]{ms};\unit[3]{ms};\unit[5]{ms};\unit[7]{ms};\unit[10]{ms}}$)
to investigate how sensitive performance is to changes in model
parameters.

Spike encoding of the preprocessed signal was modeled as
parsimoniously as preprocessing: The input signal was normalized so
that only waveform shape and not energy was considered.  Therefore, no
compensation for initial target range and associated spreading losses
was required. Spike times were determined by thresholding the signal.
Together with the specific lowpass filter chosen for the envelope
extraction step, this amounts to an ``integrate and fire'' model,
which is a simplification of the Hodgkin-Huxley
equations~\cite{Kistler-WM1997}. The sufficiency of this model for the
problem at hand will be justified below from the nature of the
features (\Sref{sec:code_properties}). As a second major
simplification, only spikes triggered by the initial transient, i.e.,
the ``onset response'' of a neuron will be considered. This
simplification is necessitated by the lack of relevant data on neural
refractoriness in bats.

A single spike time is obviously not sufficient for target
classification, since it would inevitably confound target range and
class. To retain the simplifications made already (only one bandpass
channel, only one spike triggered by the initial transient in each
neuron), a population of neurons with different thresholds was chosen
as a way to diversify the code according to the needs of target
classification.  The adopted model is therefore an amplitude-discrete
sampling of the ``inverse function'' (considering the lowest/earliest
branches only) of the preprocessed signal up to its maximum; the
signal beyond the maximum is ignored.\par

Feature extraction from spike times uses only time differences within
the neural response to an echo; using an external reference can
provide a range estimate, but has no immediate relevance for target
classification (An indirect influence is possible, should
classification features be range-dependent - this remains to be
explored). Neural circuitry for estimation of monaural time
differences is well established in bats, e.g., in the context of
ranging, where comparatively long time-of-flight values have been
found represented (few milliseconds to more than
\unit[10]{ms}~\cite{Kuwabara-N1993,Saitoh-I1995}). Mammals with
sufficient ear distances can determine direction-of-arrival by
binaural time differences, typically in the sub-millisecond range
($\unit[1]{ms}$ corresponds to $\sim\unit[34]{cm}$ distance already).
In bats, indications have been found that the respective neural
structures (MSO) can deal with time differences both in the
sub-millisecond range and beyond~\cite{Grothe-B2000}. However, this
was established only for sinusoidal amplitude modulation. In contrast
to this, the computational work presented here emphasizes the
importance of aperiodic, random time differences within echoes, which
can take values comparable to what is typically considered in binaural
difference evaluation as well as in ranging.\par

Specifically, the model consists of $M$ thresholds $a_m$, where
$a_n>a_m$ for $n>m$. These thresholds give rise to $M(M-1)$ possible
non-zero interspike intervals $\isi{m}{n}$ between the times of
crossing the $m$-th and the $n$-th threshold. For specification of the
model, two functions must be chosen; one for threshold placement on
the amplitude axis and one for selecting the threshold pairs for which
the $\isi{m}{n}$ are computed (i.e., the wiring of the neural
delay-lines/coincidence detectors). Unfortunately, no biological data
is available on either of these two functions. As a remedy, thresholds
were placed equidistantly at least one standard deviation of the noise
amplitude apart.  Since the signal-to-noise ratio of the experimental
setup, which was limited by the sound channel and not the electronics,
was better than $\unit[60]{dB}$ for the larger echo amplitudes
encountered, $M=1,024$ (chosen as an integer power of 2) thresholds
were employed altogether. Bats were found to have between $700$ and
$2\,160$ inner hair cells and between $13\,400$ and $55\,300$ spiral
ganglion cells for covering the entire hearing range of the respective
species; divergence ratios from inner hair cells to spiral ganglion
cells range from 11 to 79~\cite{Vater-M1988}. Since it is not known
how many neighboring channels could be pooled based on the similarity
of their transfer functions, it is likewise hard to estimate how many
neurons would be available for thresholding the output of one bandpass
channel. From the numbers given and the similar constraints on the
signal to noise ratio in the sound channel, it is unlikely though,
that this model sacrifices any amplitude resolution that bats may
have.\par

Once thresholds have been placed (a vector of threshold values has
been chosen), the matrix of all possible $\isi{m}{n}$ for any echo is
completely determined as well. Since this matrix has odd symmetry,
i.e., $\isi{m}{n}=-\isi{n}{m}$, considering e.g., the upper triangular
part suffices. Further more, the entire matrix can be reconstructed
exactly from the elements on the first diagonal as
\begin{equation}
  \label{eq:delta_vec_matrix}
  \isi{m}{n}=\sum_{k=m}^{n-1}\isi{k}{k+1}.
\end{equation}
In this sense interspike intervals $\isi{m}{m+1}$ generated by
subsequent exceedance of neighboring thresholds $a_m,a_{m+1}$ may be
regarded as \emph{elementary intervals}. All other intervals which may
be generated in a bat's brain are just sums of these variables.
\Eref{eq:delta_vec_matrix} describes a resolution pyramid, in which
detail is lost as the diagonal under consideration is moved away from
the main diagonal.  While the matrix of all possible \isi{m}{n}-values
is completely determined by its first diagonal and hence highly
redundant, it may be perceptually relevant, if small \isi{m}{m+1} fall
below the resolution limit, but not their sums.
\begin{figure}[htbp]
  \begin{center}
    \epsfbox{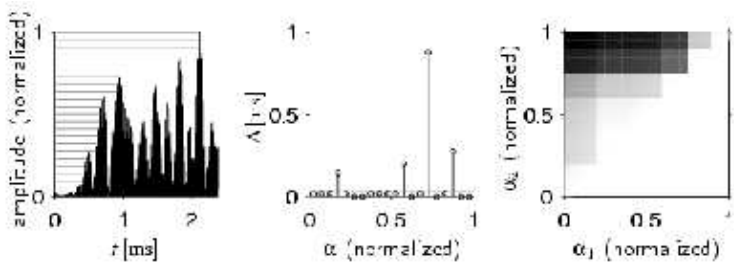}
  \end{center}
  \caption{\label{fig:model_example}Example for the application of the
    employed spike generation model. Left: the normalized waveform is
    thresholded up the its maximum; center: duration of interspike
    intervals \isi{m}{m+1} for neighboring thresholds; right: matrix
    of all possible interspike intervals for the given set of
    thresholds.}
\end{figure}\par

\section{Code properties}\label{sec:code_properties}

\subsection{Elementary interspike intervals}

\begin{figure}[htbp]
  \begin{center}
    \epsfbox{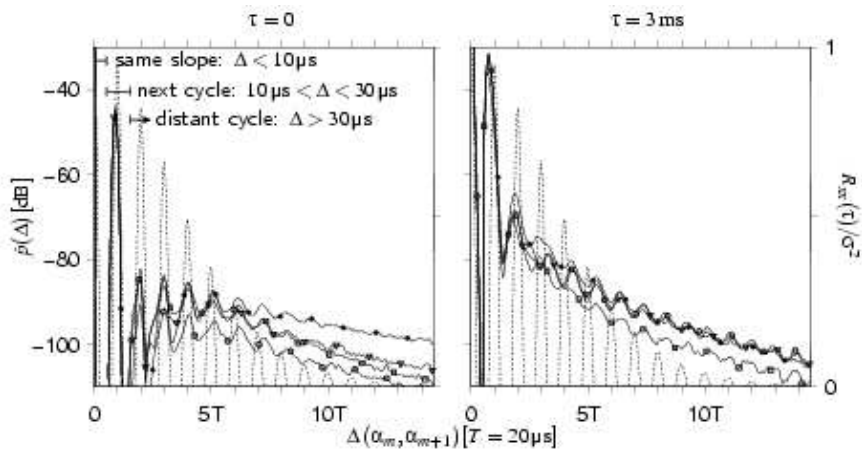}
  \end{center}
  \caption{\label{fig:dt_pdf}Probability density function estimates
    for the elementary interspike intervals \isi{m}{m+1} for the four
    foliage types sycamore (\opencircle), linden (\opensquare), maple
    (\opentriangledown), hornbeam ($\ast$). Shown are kernel density
    estimates using a normal kernel with $\unit[1]{\textrm{\textmu}s}$
    smoothing bandwidth.  Dashed line: normalized autocorrelation
    function $R_{xx}(\tau)/\sigma^2$ of the chirplet shown in
    \fref{fig:chirplet}.}
\end{figure}
Filtering the reflector sequence with the chirplet representing all
linear channel effects (\Fref{fig:chirplet}) introduces a
prominent periodicity corresponding to the carrier period of the
auditory bandpass model (here $T=\unit[20]{\textrm{\textmu}s}$). This
periodicity is clearly visible in the probability density function of
the elementary interspike intervals \isi{m}{m+1}
(\Fref{fig:dt_pdf}). Since the probability density function has
two nulls at $~\sim\unit[10]{\textrm{\textmu}s}$ and
$~\sim\unit[30]{\textrm{\textmu}s}$ for $\tau=0$ and its most pronounced
notches are in the same places for $\tau=\unit[3]{ms}$, a clear
distinction can be made between three different types of interspike
intervals depending on how the two delimiting spike times are
arranged with respect to the carrier period (\Fref{fig:dt_pdf}):
\begin{itemize}
\item For \emph{same slope intervals} flanking spikes are triggered by
  the same rising flank of a positive half-wave; for the particular
  channel center frequency chosen here,
  $\isi{m}{m+1}\le\unit[10]{\textrm{\textmu}s}$ in this interval
  category.
\item For \emph{next cycle intervals} flanking spikes are triggered by
  subsequent positive half-waves.  For the particular channel center
  frequency chosen here, next cycle intervals must have values such
  that
  $\unit[10]{\textrm{\textmu}s}\le\isi{m}{m+1}\le\unit[30]{\textrm{\textmu}s}$.
\item For \emph{distant ($>1$) cycle intervals} flanking spikes are
  triggered more than one carrier cycle apart, hence
  $\isi{m}{n}\ge\unit[30]{\textrm{\textmu}s}$ for distant cycle
  intervals.
\end{itemize}
For next and distant cycle \isi{m}{m+1}, the inverse function of the
waveform (counting only lower branches, see \sref{sec:model} and
\fref{fig:model_example}) has discontinuities, i.e.,
\begin{equation}
  \label{eq:inverse_fun_discontinuity}
  \lim_{\abs{\alpha_n-\alpha_m}\rightarrow 0}{\isi{m}{n}}
  =C
\end{equation}
as long as the discontinuity of the inverse function remains bracketed
by $\brks{\alpha_m,\alpha_n}$. This implies that such discontinuities
remain visible in any \isi{m}{n} where the corresponding thresholds
$\alpha_m,\alpha_n$ bracket them. Because they are
discontinuity-based, next and distant cycle \isi{m}{m+1} are invariant
under any monotonic non-linear transform of the signal amplitude, an
important property as the auditory system is known to perform
non-linear compression~\cite{Dau-T1996}. Unlike same slope and next
cycle \isi{m}{m+1}, the durations of distant cycle \isi{m}{m+1} are
not strictly tied to the carrier cycle, because the autocorrelation of
the chirplet ($R_{xx}(\tau)$, superposed in \fref{fig:dt_pdf})
decays and the echo waveform decorrelates.\par

\begin{figure}[htbp]
  \begin{center}
    \epsfbox{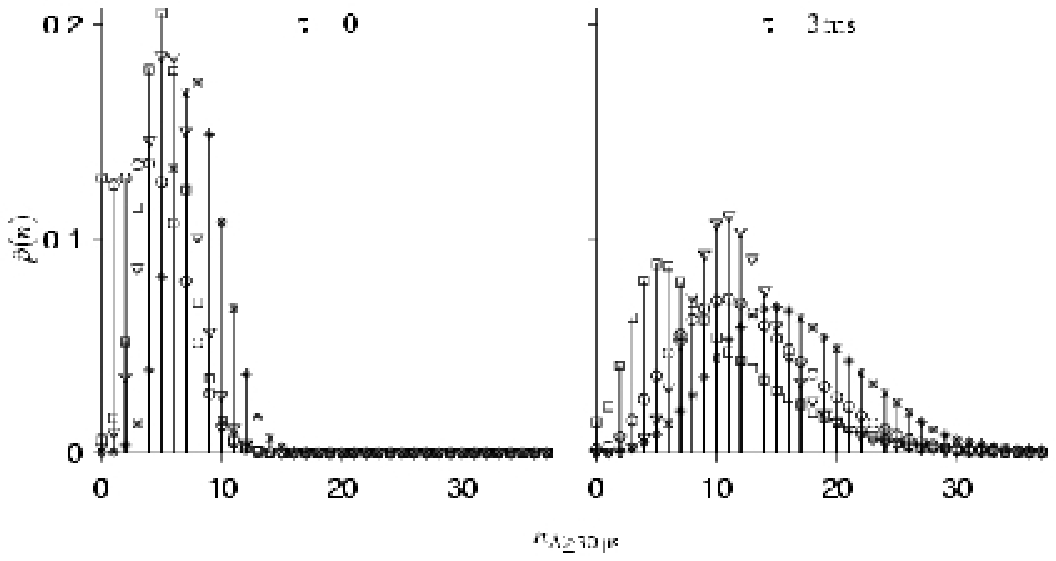}
  \end{center}
\caption{\label{fig:n_sel_pdf}Probability density function estimates
  for the number of distant cycle
  $\isi{m}{m+1}\ge\unit[30]{\textrm{\textmu}s}$ per echo. The
  probabilities for at least one distant cycle interspike interval
  $P\brcs{n_{\Delta\ge\unit[30]{\textrm{\textmu}s}}\ge1}$ is $\sim
  0.94$ for sycamore and $\tau=0$, for all others
  $P\brcs{n_{\Delta\ge\unit[30]{\textrm{\textmu}s}}\ge1}\ge 0.99$.
  Estimates used a normal kernel with smoothing bandwidth $0.82-0.99$,
  AMISE $<0.0015$. See \fref{fig:dt_pdf} for symbols denoting target
  class.}
\end{figure}

\begin{figure}[htbp]
  \begin{center}
    \epsfbox{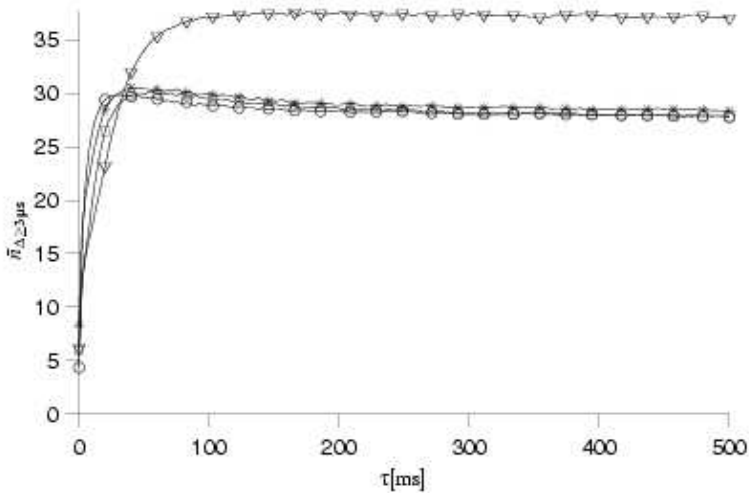}
  \end{center}
  \caption{\label{fig:lp_effect}Effect of lowpass filtering time
    constant ($\tau$) on the expected number
    $\bar{n}_{\Delta\ge\unit[3]{\textrm{\textmu}s}}$ of distant cycle
    \isi{m}{m+1}. The estimates are based on $N=500$ randomly chosen
    echoes for each value of $\tau$ and each foliage class. See
    \fref{fig:dt_pdf} for symbols denoting target class.}
\end{figure}

Despite the comparative rarity of distant cycle \isi{m}{m+1} evident
from \fref{fig:dt_pdf}, it is almost certain that at least one
distant cycle \isi{m}{m+1} is present in the response to any given
echo (\Fref{fig:n_sel_pdf}).
For the chosen threshold spacing, this is true for any smoothing
constant and the expected number of distant cycle thresholds shows a
saturating increase with increasing time constant $\tau$
(\Fref{fig:lp_effect}).
If instead the limit of $\tau\rightarrow\infty$ \emph{and}
$\abs{\alpha_n-\alpha_m}\rightarrow 0$, i.e., perfect, ``non-leaky''
integration and infinitely narrow spacing of thresholds, was to be
considered, only next cycle elementary \isi{m}{m+1} would be retained,
because they correspond to the negative half-cycles of the waveform
which were set to zero by the half-wave rectification. For finite
threshold spacing, the situation is quite different and distant cycle
\isi{m}{m+1} do not disappear as a consequence of smoothing. Both
robustness and relative rarity of distant cycle \isi{m}{m+1} are due
to the fact that these \isi{m}{m+1} are indicative of an extended
trough in the echo waveform. For any given echo, there will be but a
few of such troughs (hence the rarity), but since they are
low-frequency phenomena, they are robust against lowpass smoothing.

\subsection{Compound interspike intervals}

\begin{figure}[htbp]
  \begin{center}
    \epsfbox{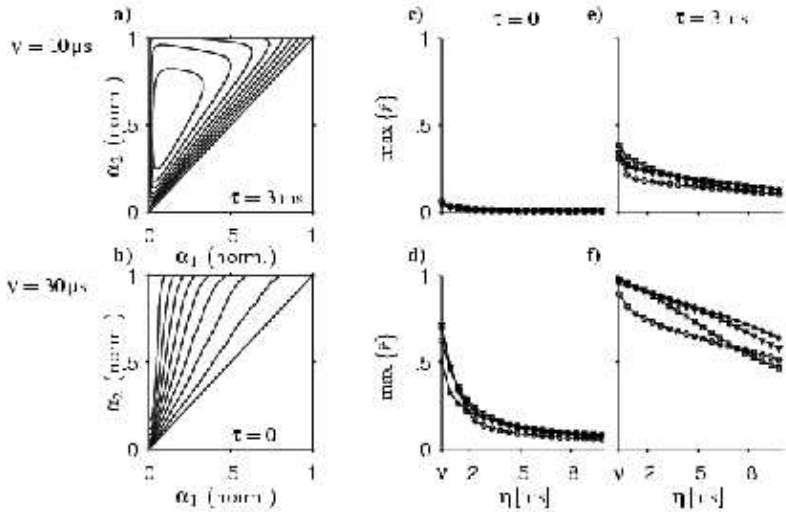}
  \end{center}
  \caption{\label{fig:r_ssb_max_of_mn}Ratio $r$ of durations of
    elementary $\isi{m}{m+1}\le\nu$ in compound \isi{m}{n} to the
    total duration of the compound $\isi{m}{n}\ge\eta$ (see
    \eref{eq:ratio_def_compute_ratio}).  Left graphs: examples of
    expected values of $r$ as a function of threshold locations
    $\alpha_1,\alpha_2$, a) $\nu=\unit[10]{\textrm{\textmu}s}$ (same
    slope intervals) and $\tau=\unit[3]{ms}$, b)
    $\nu=\unit[30]{\textrm{\textmu}s}$ (same slope and next cycle
    intervals) and $\tau=\unit[3]{ms}$. Center and right graphs:
    maximum of $r$ over all threshold pairs: c)
    $\nu=\unit[10]{\textrm{\textmu}s},\tau=0$, d)
    $\nu=\unit[30]{\textrm{\textmu}s},\tau=0$, e)
    $\nu=\unit[10]{\textrm{\textmu}s},\tau=\unit[3]{ms}$, f)
    $\nu=\unit[30]{\textrm{\textmu}s},\tau=\unit[3]{ms}$. See
    \fref{fig:dt_pdf} for symbols denoting target class.}
\end{figure}

Usage of distant cycle \isi{m}{m+1} does not place demanding
constraints on spike time resolution, whereas access to the shorter
individual next cycle intervals and in particular same slope intervals
does. Neurophysiological data on spike timing accuracy in the auditory
nerve of bats appears to be lacking, however. In cats, minimum
standard deviations of onset spike responses were found to be not much
lower than $\sim\unit[100]{\textrm{\textmu}s}$~\cite{Heil-P1997},
making individual \isi{m}{n} from the same slope and next cycle class
appear an unlikely substrate for target class estimation. Such small,
elementary \isi{m}{m+1} could achieve guaranteed perceptual saliency,
however, if the amplitude range spanned by a threshold pair was
widened. In this case, longer, resolvable compound intervals (see
\eref{eq:delta_vec_matrix}) could emerge as a sum of elementary
\isi{m}{m+1} with the value of the sum being dominated by
contributions from same slope and next cycle \isi{m}{m+1}. To explore
this possibility, compound \isi{m}{n} which exceeded some minimum
length $\eta$
\begin{equation}
  \label{eq:ratio_def_keep_sum}
  \isi{m}{n}=\sum_{k=m}^{n-1}\isi{k}{k+1}\ge\eta
\end{equation}
were selected and a ratio $r$ which describes the contribution of
elementary $\isi{m}{m+1}\le\nu$ (i.e., same slope or same slope or
next cycle intervals) was computed as
\begin{equation}
  \label{eq:ratio_def_compute_ratio}
  r=\frac{\sum_{k=m}^{n-1}\isi{k}{k+1}I_{\isi{k}{k+1}\le\nu}}
  {\sum_{k=m}^{n-1}\isi{k}{k+1}},
\end{equation}
where $I$ is the indicator function. The expected value of this ratio
was found to depend on the choice of $m$ and $n>m$ (see examples in the
left subgraphs in \fref{fig:r_ssb_max_of_mn}) and therefore the
expected overall impact of same slope and next cycle interspike
intervals on the interspike intervals actually read out cannot be
estimated without knowing the distribution of readout connections over
all possible pairs of incoming neurons. The maximum ratio is a
distribution-free measure, however, and it indicated that same slope
\isi{m}{m+1} have little impact on long compound \isi{m}{n} regardless
of the smoothing time constants (\Fref{fig:r_ssb_max_of_mn}).
Next cycle elementary \isi{m}{m+1} could be the dominating component
of long compound \isi{m}{n}, if a long smoothing time constant was
chosen. On the basis of these results, same slope elementary
\isi{m}{m+1} are of doubtful perceptual salience, both in isolation
and in compound intervals. Next cycle elementary \isi{m}{m+1} are of
doubtful perceptual salience in isolation, but may be a dominating
component in longer compound \isi{m}{n}, if long integration times are
chosen.  Therefore, in the next section (\Sref{sec:isi_rp}), only
next and distant cycle \isi{m}{m+1} are retained for further
consideration.

\subsection{Interspike interval random process}\label{sec:isi_rp}

Retaining only next and distant cycle \isi{m}{m+1}, each echo is
represented by a random sequence of variable length, since for each
\isi{m}{m+1}-class more than one \isi{m}{m+1} per echo is likely (see
Figures~\ref{fig:n_sel_pdf},\ref{fig:lp_effect} for distant cycle
\isi{m}{m+1}; next cycle \isi{m}{m+1} are more common than distant
cycle \isi{m}{m+1}, see \fref{fig:dt_pdf}). Associated with each
\isi{m}{m+1} is a position along the amplitude axis marking the
location of the two neighboring thresholds the flanking spikes were
triggered at.\par

\begin{figure}[htbp]
  \begin{center}
    \epsfbox{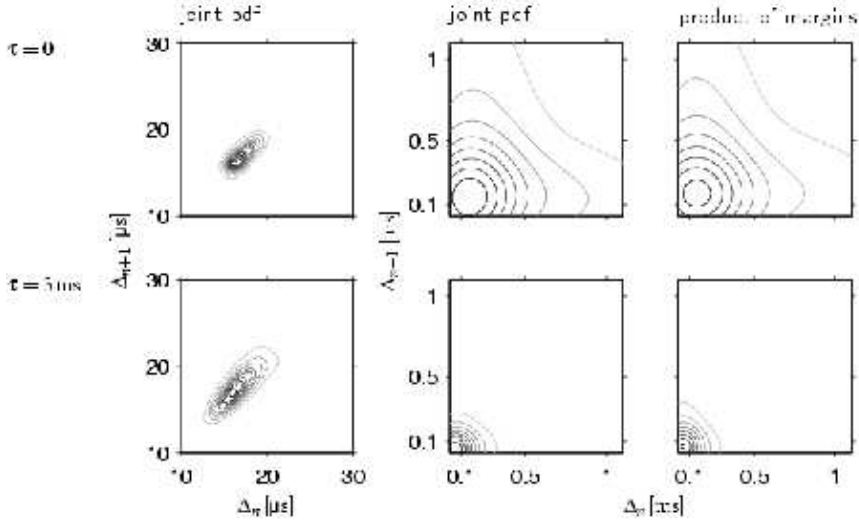}
  \end{center}
  \caption{\label{fig:dt_pair_indep_all_ex}Estimates of the joint
    probability density function for neighboring \isi{m}{m+1} in the
    response to echoes from sycamore foliage. The contour levels are
    spaced linearly between 10\% and 90\% of the density functions'
    maxima. Estimates are based on $N=21\,200$ echoes.}
\end{figure}

\begin{figure}[htbp]
  \begin{center}
    \epsfbox{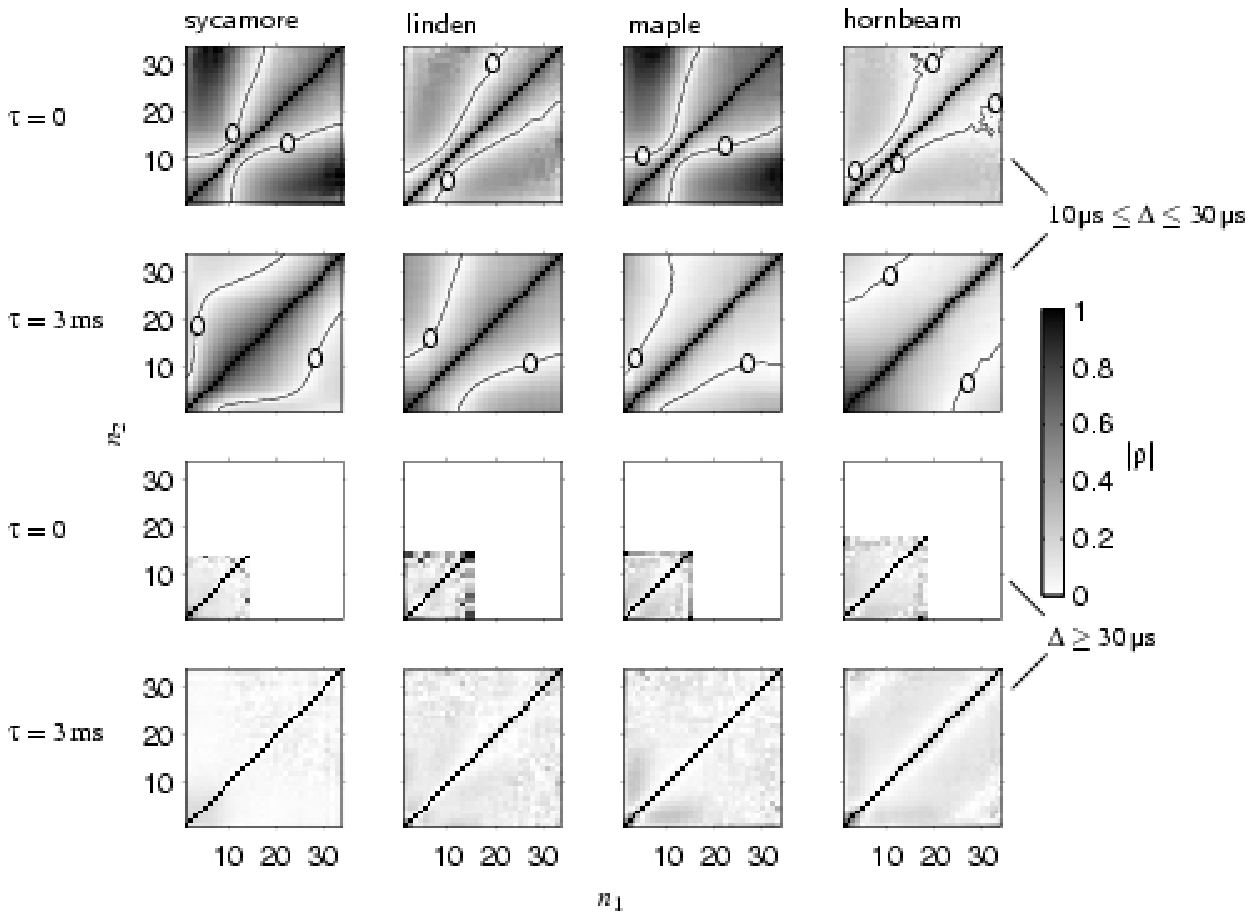}
  \end{center}
  \caption{\label{fig:dt_cov_ex}Correlation matrix estimates for the vectors
    of next cycle and distant cycle \isi{m}{m+1}. The matrices show
    estimates of the correlation coefficient magnitude $\abs{\rho}$
    for interval duration as a function of positions $n_1,n_2$ in the
    interval sequence. The top two rows show correlation matrix
    estimates for next cycle \isi{m}{m+1} generated with smoothing
    time constants $\tau=0$ (first row) and $\tau=\unit[3]{ms}$
    (second row). The bottom two rows show the same estimates for
    distant cycle \isi{m}{m+1}.  Estimates are based on $N=21\,200$
    echoes for each foliage class; the number of realizations for
    sequences of a particular minimum-length varies, however, and
    leads to a higher estimator variance on the edges of the
    covariance matrices for $\Delta\ge\unit[30]{\textrm{\textmu}s}$
    and $\tau=0$.}
\end{figure}

The random sequences formed by next and distant cycle \isi{m}{m+1}
differ in their statistical properties: Next cycle \isi{m}{m+1} show a
strong pairwise dependence between neighboring values as well as
correlations of varying strength over the entire sequence, whereas
distant cycle \isi{m}{m+1} random sequences are uncorrelated and at
least pairwise independent
(Figures~\ref{fig:dt_pair_indep_all_ex},\ref{fig:dt_cov_ex}).
Therefore, the distant cycle \isi{m}{m+1} random sequences have a much
simpler statistical structure than next cycle \isi{m}{m+1}, which
facilitates the design of a classifier. For this reason, as well as
because of their low resolution requirements, they will be used in the
next section to attempt target classification based on output of the
spike generation model (see \sref{sec:classification}).

\section{Classification based on distant-cycle interspike
  intervals}\label{sec:classification}

The results outlined in the previous sections demonstrate that
distant-cycle $\isi{m}{m+1}$ offer advantages both for actual use by
biological systems (low resolution requirements, high visibility in
compound $\isi{m}{n}$) as well as for further studies (uncorrelated
random sequences). The decisive question is whether the distant-cycle
$\isi{m}{m+1}$ also contain sufficient information on target class. To
answer this question, target classification was attempted using an
ad-hoc feature selection approach, which is unlikely to make optimum
use of the random sequences, but serves its purpose of demonstrating
feasibility in case of success.\par

Each spike response to an echo was represented by three features:
first moment estimates for distant cycle $\isi{m}{m+1}$ interval
length ($\bar{\Delta}$) and amplitude location ($\bar{\alpha}$) as
well as the number ($n_{\Delta\ge\unit[30]{\textrm{\textmu}s}}$) of
distant cycle $\isi{m}{m+1}$ in the spike response to an echo:
\begin{equation}
  \eqalign{
  \label{eq:features}
  n_{\Delta\ge\unit[30]{\textrm{\textmu}s}}
  &= \sum_mI_{\isi{m}{m+1}\ge\unit[30]{\textrm{\textmu}s}}\\
  \bar{\alpha} 
  &=\frac{1}{2N}\sum_m\prns{\alpha_{m+1}-\alpha_m}
  I_{\isi{m}{m+1}\ge\unit[30]{\textrm{\textmu}s}}\\
  \bar{\Delta}
  &=\frac{1}{N}\sum_m \isi{m}{m+1} I_{\isi{m}{m+1}\ge\unit[30]{\textrm{\textmu}s}}}.
\end{equation}
While not providing a sufficient statistic, settling for first moments
is well advised in the light of the small sample nature of the
obtained spike representation (see \fref{fig:n_sel_pdf}): Since
both, $\alpha_{m+1}-\alpha_m$ and \isi{m}{m+1} are positive
quantities, estimates of first moments are more robust than those for
all higher moments (at least if a sample average estimator or
equivalent is used~\cite{Bourin-C1998}). A biological implementation
of this feature space is also readily envisioned, e.g., the center of
gravity of the excitation on neural maps for amplitude and time delay
would represent $\bar{\Delta}$ and $\bar{\alpha}$, the total
amount of excitation $n_{\Delta\ge\unit[30]{\textrm{\textmu}s}}$.\par

\begin{figure}[htbp]
  \begin{center}
    \epsfbox{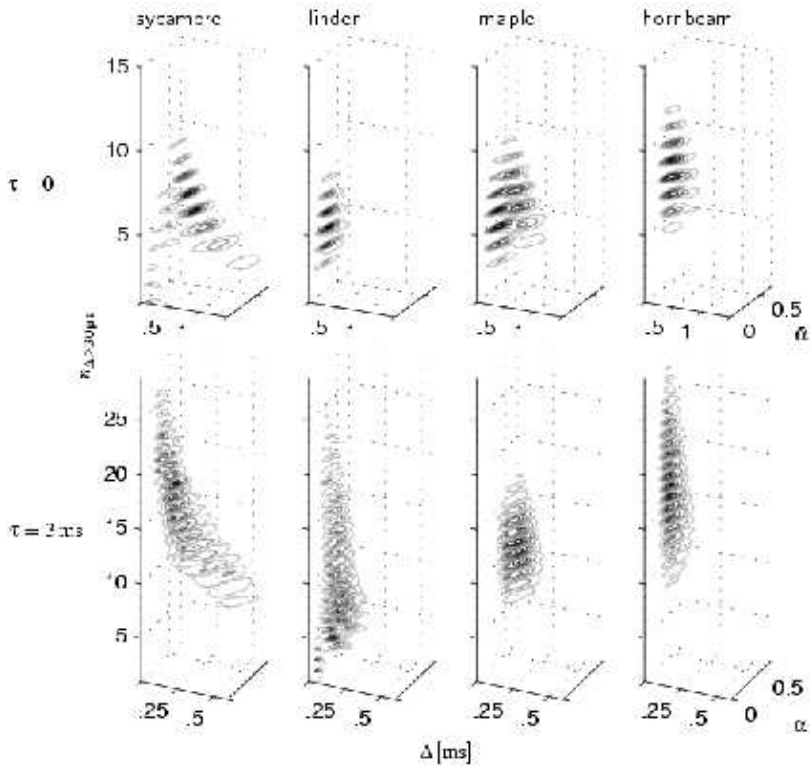}
  \end{center}
  \caption{\label{fig:dt_mom_a_mom_n_sel_jpdf_cslice}Estimates of
    joint probability density functions for the code features
    $\bar{\Delta}, \bar{\alpha},
    n_{\Delta\ge\unit[30]{\textrm{\textmu}s}}$ (see
    \eref{eq:features}). Top row: $\tau=0$; bottom row:
    $\tau=\unit[3]{ms}$. Estimates are based on $N=21\,200$ echoes
    each.}
\end{figure}

The three-dimensional joint probability density functions
(\Fref{fig:dt_mom_a_mom_n_sel_jpdf_cslice}) of the features (see
\eref{eq:features}) show interesting structure (e.g., multimodality
for maple echoes) as well as dependencies between the features (e.g.,
for sycamore echoes, there tend to be either few large or many small
distant cycle \isi{m}{m+1}).
The suitability of the distances between the probability density
functions for target classification was assessed by estimating
performance measures of an m-ary sequential probability ratio
test~\cite{Baum-C1994}. Because bats use pulse trains with repetition
rates that are typically high compared to the time scales that
navigation decisions are made on, this approach provides the necessary
model to explain how bats could make use of the information which
accumulates over the incoming echo trains.\par

\begin{figure}[htbp]
  \begin{center}
    \epsfbox{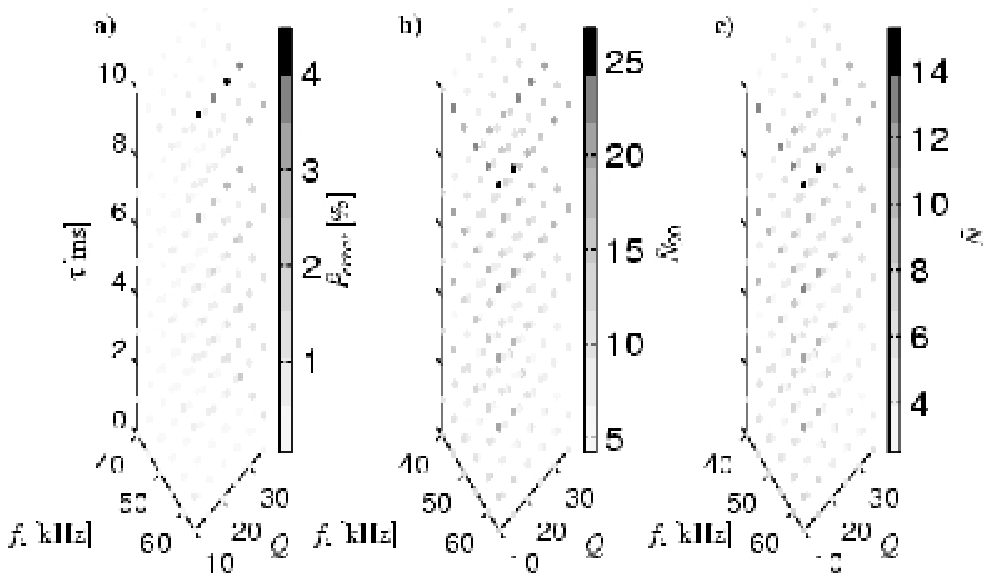}
  \end{center}
  \caption{\label{fig:dt_mom_a_mom_n_sel_msprt_par}Dependence of
    classification performance on preprocessing model parameters.
    Parameters are: center frequency of the auditory bandpass channel
    model $f_c$, its $-\unit[3]{dB}$ filter quality $Q$, and smoothing
    time constant $\tau$. Performance measures are: a) Estimated
    classification error probability $\hat{P}_{\textrm{error}}$
    (averaged over target class), b) 90\%-percentile ($\hat{N}_{90}$)
    and c) expected value ($\bar{N}$) for the number of samples
    (echoes) needed for a decision.}
\end{figure}

The classification trials were conducted based on random draws of
echoes from the stimulus ensemble, this discards any information which
may be provided by systematic changes in echo features over a certain
path~\cite{Kuc-R2001}. Nevertheless, an excellent classification
performance was found (\Fref{fig:dt_mom_a_mom_n_sel_msprt_sum}):
\begin{figure}[htbp]
  \begin{center}
    \epsfbox{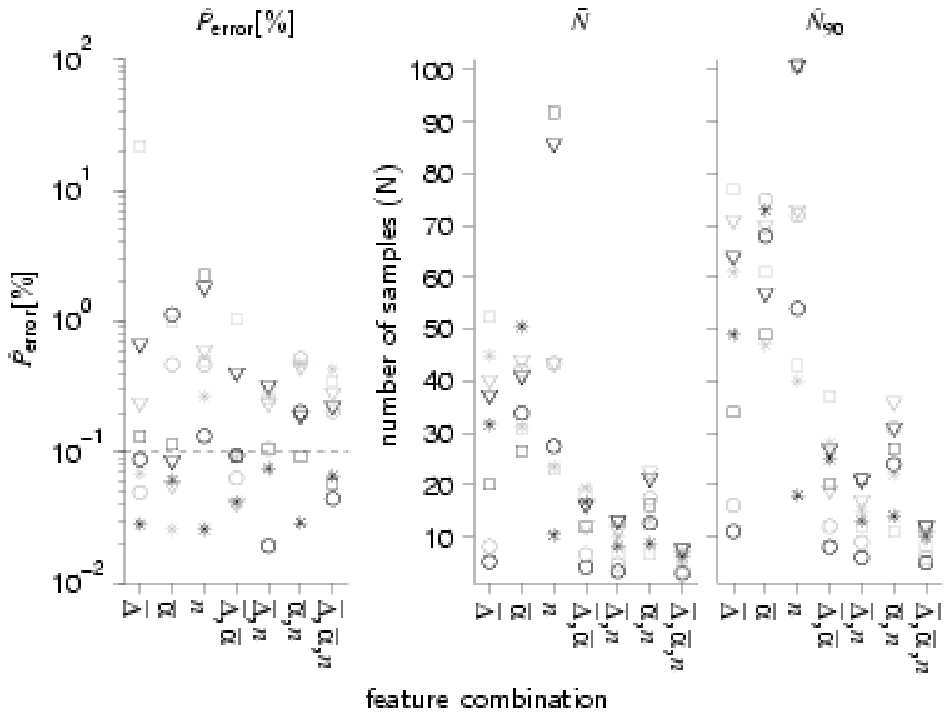}
  \end{center}
  \caption{\label{fig:dt_mom_a_mom_n_sel_msprt_sum}Classification
    trial results for the four foliage types based on the three
    distant cycle \isi{m}{m+1} features ($\bar{\Delta}, \bar{\alpha},
    n_{\Delta\ge\unit[30]{\textrm{\textmu}s}}$, see
    \eref{eq:features}) and their combinations. Left: estimated class
    conditional error probability \unit{\lbrack\%\rbrack}; center:
    expected number of samples (echoes) needed for a decision; right:
    90\%-percentile of the number of samples. Black symbols: $\tau=0$;
    gray symbols: $\tau=\unit[3]{ms}$. Responses of the spike coding
    model were drawn randomly from $21\,200$ examples for each class,
    $N=10^5$ trials were conducted for each performance estimate. See
    \fref{fig:dt_pdf} for symbols denoting target class.}
\end{figure}
Using the joint probability density function of all three features and
no smoothing ($\tau=0$), error probabilities of 0.03 to 0.19\% were
obtained on an expected number of 3 to 8 echoes (90\%-percentiles
ranged from 6 to 13). For moderate smoothing ($\tau=\unit[3]{ms}$), a
slight performance decrease was found (error probabilities: 0.24 to
0.5\%, expected number of samples: 5 to 8, 90\%-percentiles: 9 to 13,
see \fref{fig:dt_mom_a_mom_n_sel_msprt_sum}). Using the joint
probability density function of all three features was found to result
in the best overall performance, so both first order properties of the
neural response as well as the number of time intervals contain target
class information. Using all three features, the overall dependence of
classification performance on preprocessing model parameters ($f_c$,
$Q$, $\tau$) was found to be weak, average values (over all four
target classes) of error probabilities, sample numbers and their
90\%-percentiles were found to fall in the intervals 0.06-4.5\%, 3-15
and 4-27 respectively. The least favorable values were outliers which
were reached for a few, adverse parameter combinations only
(\Fref{fig:dt_mom_a_mom_n_sel_msprt_par}).
These results demonstrate that the parameters of the preprocessing
model are of little relevance within the parameter ranges
($f_c=\brks{\unit[40]{kHz},\unit[65]{kHz}},
Q=\brks{10,35},\tau=\brks{0,\unit[10]{ms}}$) studied.\par

\section{Conclusions and directions for future research}

The present work addresses the acoustic landmark identification as an
example problem of biomimetic random process classification. Because
the sensory representation of sound is low-dimensional, most biosonar
sensing tasks involving extended, multi-faceted targets are likely to
be posed in the way of random process estimation problems. In vision,
this situation is much less common, because retinal images leave fewer
alternative interpretations and often additional assumptions are
available to decide between them. This leads to regularization
approaches being considered as models of visual
perception~\cite{Poggio-T1985}, which would fail in biosonar
perception. The specific merit of biosonar as a sensory model system
lies therefore in the fact that it matches vision in sustaining
animals with active mobility in three-dimensional space despite this
severe ill-posedness.\par

For the studied example problem, possible solutions were explored in a
computational approach on a spike code level. The use of a
parsimonious model for generating this spike code aides the search for
basic, robust principles. Highly informative and accessible code
features should be readily visible in the output of any model which
reproduces the relevant principles correctly. The basic assumptions
made here were the well-established view that spike-generation can be
approximated as smoothing followed by thresholding and that
time-differences are the elements of the code.  The latter assumption
is particularly appealing in bats, where small, monaural time
differences are known to be behaviorally relevant as well as neurally
extracted. In principle, however, the discovered features (extended
troughs in the waveform) may as well be accessible in other codes,
like e.g., a rate code. In bats, a rate code would have to be
reconciled with the fact that signals of large bandwidth must be coded
with a comparatively small number of auditory nerve fibers, which may
result in excessively large estimator
variance~\cite{Gautrais-J1998}.\par

The interval code served as a biomimetic guide for identifying
classification features. The central insight gained is that within all
possible elementary interspike intervals (formed between neighboring
thresholds) which an echo can generate, a few, comparatively long
distant-cycle intervals stand out: They are readily resolved in
isolation already and furthermore are the dominating component in any
compound interspike interval (formed between distant thresholds) they
are part of. Distant-cycle interspike intervals can be viewed as an
acoustic analogue to edges in a visual image: They are the result of a
discontinuity (in the inverse function of the waveform in the acoustic
case) and are readily visible over a range of different resolutions
(i.e., amplitude threshold spacings in the acoustic case). However,
whereas in visual images edges tend to delineate the shape of
deterministic objects or patterns, for echoes this not the case.
Therefore, the problem of dealing with ``echo edges'' is not a pattern
recognition problem, but a random process classification problem
without a deterministic template.\par

The chosen example problem (classification of different foliages)
holds little promise for classic feature selection methods: the
probability density functions of signal amplitude are
non-Gaussian~\cite{Mueller-R2000a}, and the only non-negligible
structure in the auto-covariance matrix is determined by the sonar
pulse. Nevertheless the number, average duration and average amplitude
location of the few distant-cycle interspike intervals in the spike
response to each echo class were shown to provide excellent target
class information. Therefore, the features which were found to be of
high visibility in the spike code derived from a parsimonious model
also proved to be highly informative.\par

Further work is needed to elucidate the structural basis of these
features, i.e., what kind of physical target properties they
correspond to. These could be the distributions of individual
reflector properties (e.g., size, spatial orientation), properties of
their spatial distribution or, more specifically, properties of the
contours that limit these spatial distributions. In this way, the
findings for the example stimulus ensemble considered here could be
generalized to a more inclusive theory about the information that is
accessible to biosonar systems in natural environments.\par

Finally, the coding model investigated has been limited to isolated
portions (a single auditory bandpass channel) of the auditory signal
representation and to random sequences of echoes. Relationships which
may exist across the frequency dimension of the auditory signal
representation~\cite{Smith-1996} or across the echo
sequence~\cite{Kuc-R2001} generated along a particular flight path of
a bat have been ignored. In the view of these omissions, the achieved
classification performance is particularly remarkable. Using the full
information available across frequency and scan path, bats may be able
to make even finer discriminations (e.g., identifying different trees
of the same species, different views or portions of the same tree).
Spatial gradients explored along a flight path could be used for
performing estimation tasks other than target classification, for
instance, path planning, e.g., in the form of contour following, could
be performed by following a spatial gradient in statistical echo
properties.  Assuming that the nature of such spatial gradients would
depend on target class, research into the existence and information
contend of spatial in the studied features would link target
classification to a much wider set of tasks that animals need to
perform in their natural habitats.

\ack
Supported by DFG (SFB 550, project B6).

\end{document}